# Enhancing Video Transmission with Machine Learning based Routing in Software-Defined Networks

Anıl Dursun İpek[1], Murtaza Cicioğlu[1*], Ali Çalhan[2]

[1]Department of Computer Engineering, Bursa Uludağ University, 16059, Bursa, Türkiye
[2]Department of Computer Engineering, Düzce University, 81620, Düzce, Türkiye
*Corresponding author: murtazacicioglu@uludag.edu.tr

**Abstract**

In traditional networks, the main problems are the lack of network management and a centralized approach. Additionally, many vendors use non-standardized devices resulting in a complex and difficult-to-manage network structure. Our study uses the centralized, flexible, dynamic, and programmable structure of Software-Defined networks (SDN) to overcome the problems. Although SDN effectively addresses the challenges present in traditional networks, it still requires further enhancements to achieve a more optimized network architecture. The Floodlight controller utilized in this study employs metrics such as hop count, which provides limited information for routing. In scenarios such as video transmission, this situation is insufficient and the need for optimization arises. For this purpose, an artificial intelligence (AI) based routing algorithm is proposed between the server and the client in the scenario based on NSFNET topology. The topology designed with the Floodlight controller in the Mininet simulation environment includes a client, a server, and 14 switches. A realistic network environment is provided by adding different receivers and creating TCP traffic between these receivers using the iperf3 tool.

In three scenarios, video streaming is performed using the FFmpeg tool, and 49 path metrics such as RTT, throughput, and loss are recorded. In these scenarios, PSNR and SSIM calculations are made to observe the differences between the transmitted and the original video in congested and uncongested environments. Due to the lack of a dataset suitable for the proposed network environment in the literature, a new dataset consisting of 876 records is created using continuously transmitted video traffic. When the obtained dataset was examined, it was determined that 19 out of 49 features affected the traffic level. Low and high traffic levels are created within the dataset, and different machine learning techniques such as KNN, Random Forest, SVM, AdaBoost, Logistic Regression and XGBoost are applied using the features that affect the traffic levels.

As a result, machine learning models successfully classify the traffic level. The Logistic Regression model that performs best, is used in the routing algorithm and selects the route to be transferred according to the traffic level before video transmission. The proposed AI-based routing algorithm is more successful when comparing the Floodlight controller routing algorithm with a hop-count metric. With the proposed approach, traffic on the network can be handled and detected from different dimensions with many parameters.

**Keywords:** SDN, Machine Learning, Routing, QoS.

1. **Introduction**

The use of computer networks is becoming increasingly common as the number of devices on the network increases exponentially. According to the Cisco Annual Internet Report, the number of network devices, which was 18.4 billion in 2018, is expected to reach 29.3 billion by the end of 2023 [1]. As a result of these developments, traffic on the network is gradually increasing and this trend is expected to continue with further technological developments. This increase in connected devices poses significant challenges in managing traditional network systems that often rely on built-in management protocols within each device [2–4].

Network devices operate on two distinct planes: the data plane and the control plane. The network nodes make routing decisions and create communication paths for data packets in the control plane. The data plane is responsible for physically transmitting these packets over the network. The main challenge in traditional network systems is that each network device makes and implements its own routing decisions independently of the others. The lack of centralized control, compounded by the proprietary protocols employed by different vendors, hinders the development of a universal communication language. Consequently, integrating new functionalities and network updates becomes a complex and vendor-specific task, ultimately increasing management costs [5–8].

The idea of software-defined networks (SDN) was put forward in studies conducted at Stanford University in 2008 to solve the problems of existing traditional networks. The basic principle of these networks is to address the scalability, management, flexibility, and cost problems in traditional networks and to bring an innovative approach. In SDN architecture, control, and data planes are separated for creating a centralized approach. The controller in the control plane is the brain of the network and is responsible for managing the network. Controllers are computers with high processing power and make general decisions on the network. Switches, routers, and other intermediate devices (middlebox) are in the data plane and they implement the decisions taken in the control plane and are only physically involved in packet transmission [9–11].

The emergence of SDN has revolutionized network management by enabling a holistic view of the network topology. This paradigm shift empowers network programmability, fostering increased efficiency and optimization. Improving resource utilization and increasing user service quality are needs of today's networks. When the service quality of traditional networks is examined in areas such as data centers and content distribution networks, it is seen that they are inadequate due to their lack of a global view. While traditional networks employ load balancing and routing protocols to manage the quality of service and optimize network performance, research has demonstrated that these methods are demonstrably less efficient comparing their SDN counterparts [10,12–15].

The load balancing and routing algorithms have similar functions. They have basic tasks such as minimizing response time on the network, optimizing resource usage, avoiding bottlenecks, and maximizing efficiency. This study proposes an AI-based routing algorithm for data transfer between the server and the client in a virtual environment. In the simulation environment, the controller selects the data transmission path by making decisions based on limited information, typically using a single parameter such as hop count or latency during the routing. More reliable data transmission is made possible by the controller's centralized function, which enables the acquisition of certain parameters during data transmission between two network devices. However, as the number of parameters used in the routing increases, the decision-making mechanisms must also be optimized. In this study, a multi-parameter AI-based algorithm is proposed to optimize the routing process for data transmission between the server and client by leveraging the controller's powerful capabilities. The flexible and centralized structure of SDN is used in the design of the algorithm. It is done by dynamically evaluating the general view of the network with AI models. This study aims to reduce network traffic, optimize bandwidth usage, and improve user experience.

AI is an approach that aims to provide human-like intelligence features to computer systems [16]. AI is accelerating at an unprecedented pace, powered by the exponential growth of computing power, vast datasets, and advances in subfields such as machine learning and deep learning. In network management, AI exerts a transformative influence across a diverse spectrum of applications. Smart home systems create smarter networks by communicating with automated vehicles and other Internet of Things (IoT) devices. It plays an important role in network security in detecting, analyzing, and preventing attacks [17]. AI algorithms are adept at detecting potential security breaches by analyzing network traffic for abnormal patterns such as unusual traffic volumes, unauthorized access attempts, or suspicious data packets, thereby increasing overall network resiliency. It also optimizes network performance, manages data traffic, increases transmission speed, or contributes to more efficient operation of networks by regulating energy use. Heuristic algorithms, informed by AI decision-making models, can be incorporated into routing protocols to enable dynamic traffic management on the network [18,19].

Floodlight controller and Mininet are used as the development environment in this study. Mininet is a simulation tool that provides a virtual network development environment and offers Python API support. Floodlight is an SDN controller being developed in Java. Floodlight offers a language-independent development environment between the network controller and the programmer with the REST API. The iperf3 tool is used to generate traffic during the development phase of the study. Therefore, UDP and TCP traffic can be created on the network. To measure user quality of service in the network, video is transferred with the ffmpeg tool. After the transfer, the initial version of the video is compared and as a result, PSNR (Peak Signal-to-Noise Ratio) and SSIM (Structural SIMilarity Index) values are produced. The change in video quality is evaluated with the data obtained.

The study offers several significant contributions to network management utilizing AI. Firstly, it enhances user experience by facilitating lower latency data transfers and enabling the delivery of higher-quality content, such as high-resolution video streams. Secondly, the model, trained on a comprehensive dataset specifically created within a virtual simulation environment using Mininet, dynamically optimizes network traffic by making real-time routing decisions. This approach fosters improved network performance and efficiency. Thirdly, the study promotes efficient resource utilization by leveraging AI for intelligent resource allocation. By distributing the load equally across available data paths, the model prevents unnecessary resource consumption and optimizes network operations. Finally, the study contributes to the advancement of the field by creating a novel dataset within the Mininet testbed, which can serve as a valuable resource for future research endeavors.

In addition to the main contributions, our study contains various difficulties in the implementation phase. These challenges are:

- Implementing the application within a virtual environment presented a unique challenge. Due to the absence of physical distances between network devices in Mininet, routing decisions solely based on hop counts introduced complexities in the application logic.

- The model's training relied heavily on a suitable dataset for the virtual environment. The lack of relevant datasets in existing literature necessitated the creation of a custom dataset within the Mininet testbed. This highlights the importance of domain-specific datasets for achieving optimal performance in machine learning and deep learning applications.
- While the controller efficiently handled tasks like traffic prediction and switch communication during routing, neglecting its CPU consumption could pose scalability limitations in the future. Our observations indicated a gradual increase in controller CPU utilization (mention specific percentage or processing time increase if measured). Implementing resource management strategies or exploring alternative controller architectures could be potential areas for future investigation.
- Compared to the theoretical studies, limited practical applications brought another difficulty. This gap emphasizes the need for more research efforts focused on the real-world application of AI-powered network management solutions.
- Finally, collecting network information from a limited number of simulated tools within the constraints of the simulation environment resulted in disregarding the time factor during data collection. The main reason for this problem is that the simulation environment has certain restrictions.

This study investigates the application of AI for dynamic traffic routing in SDN. We leverage the Mininet framework to emulate the NSFNET topology and implement a novel AI-based routing algorithm. This algorithm optimizes data transfer paths between designated clients and servers, prioritizing low-traffic routes to enhance user experience. The performance of our proposed approach is evaluated by comparing its throughput and latency metrics against the hop-count-based routing employed by the Floodlight controller. The results, presented graphically, demonstrate significant improvements in network performance achieved through our AI-driven routing strategy. Section 2 includes a summary of the relevant study and research conducted in the literature. In Section 3, the tools and development environment used are explained. The materials and methods are given in detail in Section 4, and finally, in Section 5, the results are presented.

2. **Related Works**

In his study, Ahmet W. [20] presents a work about the Equal-Cost Multipath (ECMP) routing algorithm, one of the classical load balancing algorithms, and the bandwidth load balancer he developed. ECMP is statically run on the network and evenly distributes incoming streams to the next network devices. However, its major disadvantage is its static nature, as it is unaware of the overall state of the network. In contrast, the bandwidth load balancer dynamically transfers streams based on bandwidth statistics received by the controller, according to the approach presented in the study. The approach allows switches to be aware of the network's general state and forward flows to directions with no traffic. The algorithms are compared in a topology created using the Floodlight controller in the Mininet environment. According to the results, the proposed approach increases network throughput by 30%. However, the proposed approach also faces some challenges. The tests are conducted on a small topology, so the algorithm needs further development to be integrated into larger networks. Additionally, the tests neglected the CPU load on the controller, which is a known issue as the processing capacity of the controller is limited and may cause delays in resource usage situations.

Bilal B. and Banu U. [21] research load balancing in data centers using AI techniques. Traditional load-balancing methods are deemed insufficient due to the intensity of data center use and the rapid increase in data volume. SDNs offer dynamic load balancing as an alternative to traditional methods in data centers. This paper focuses on utilizing AI techniques such as Artificial Neural Networks (ANN), Support Vector Machines (SVM), Machine Learning (ML), and Deep Learning (DL) for load balancing. The goal is to reduce delays between servers and enhance overall system performance. A dataset was created by conducting tests on the system topology created using SDNs. This dataset comprises 800 training data points and 192 test data points. The most successful model identified through training is ANN. Integrating AI into the system has led to a performance increase.

Ahmed H. and Ahmadreza M. [22] examine AI-based load balancers used in SDN from a broad perspective. First of all, it divides load balancers into 4 classes. The study categorizes load balancers into four classes: nature-inspired load balancers, machine learning-based load balancers, mathematical model-based load balancers, and other load balancers. Under each heading, studies conducted in the literature after 2017 were examined in detail. A comparison table is created to highlight the advantageous and disadvantageous parts of the research. The parameters used for load balancing are listed according to their frequency of use and presented as a reference for future studies. According to the research, the five most used network parameters for load balancing are throughput, response time, latency, workload degree, and packet loss ratio. The study emphasizes that previous studies did not use sufficient quality of service (QoS) parameters and provides a good perspective for future studies.

Mohamed I. Hamed et al.'s study [23] discusses the problems associated with traditional load balancers, such as lack of flexibility and difficulties in managing network flows. The article proposes the use of SDN as a solution to these problems. Traditional load balancers are inflexible and have been criticized for high hardware costs. SDN is suggested as a solution because it offers centralized control and programmability, which can address these issues. The study introduces a bandwidth-based load-balancing approach that distributes network demands among servers based on bandwidth consumption. The performance of the proposed approach is evaluated against Round Robin and Links-based load balancing schemes in the Mininet simulation environment and Raspberry Pi application. The results indicate that the proposed approach outperforms these traditional schemes. Overall, the study concludes that SDN can provide an economical and flexible solution to the problems associated with traditional load balancers.

Mosab H. and colleagues [12] wrote an extensive research paper on load-balancing techniques. They begin by discussing the advantages of SDN over traditional networks. Then, they delve into the basic objectives of load balancers, which include minimizing response time, optimizing resource usage, avoiding bottlenecks, and maximizing throughput. Load balancers are categorized as data plane and control plane load balancers. The data plane is further divided into server and connection load balancers, while the control plane is subdivided into logically centralized, physically distributed, and virtualized load balancers. Research findings indicate that SDN-based load balancers outperform traditional load balancers. Additionally, the paper summarizes the network parameters used in load-balancing methods and discusses security issues in SDN.

In the study by Mohammad R. B. et al. [18], the authors discuss how the increase in service demands from cloud users significantly raises the density and load on the network. This reduction in efficiency leads to system slowdowns. To improve system efficiency, it must be capable of handling various loads and increasing needs and show dependencies on automatic measurement systems. The research presented examines two AI optimization techniques, Ant Colony Optimization (ACO) and Particle Swarm Optimization (PSO), to investigate how these techniques can effectively solve load-balancing problems in SDN. An SDN network topology of 6 switches and 8 hosts was used in an experiment to compare the performance of ACO and PSO. Through the analysis, it is observed that PSO has lower latency and better performance. The study demonstrates a method to increase performance and improve load-balancing capabilities through modifications to existing AI optimization techniques. As a result, the study is proposed to evaluate node and link reliability in SDN.

Ekber C. K. [24] proposes an architecture with a new routing algorithm based on the user access protocol for the eMBB (Enhanced Mobile Broadband) content of the 5G network. He carries out his work using the Floodlight controller in the Mininet environment. In the SDN topology, 2 cache servers, 1 receiver, and 6 additional receivers were used to generate traffic. One of the servers is located very close to the receiver, but there is TCP or UDP traffic on the path between the receiver and the server. The other server is in a traffic-free environment, but there is a greater distance between it and the receiver. In his study, the receiver requests video content from the servers. The change in video quality is observed by comparing the transferred videos before and after transmission. The quality parameter used at this stage is PSNR. The Dijkstra algorithm that finds the shortest distance has been used. The researcher compares the algorithm that routes according to traffic density, with traditional DNS-based routing which considers the number of hops in the topology, and concludes that it gives a 60% better PSNR value.

In their study, Ayşe N. T. and Mevlüt E. [25] classify security threats occurring in SDN with machine learning algorithms. Attacks were carried out by sending DDOS attacks in the topology created on Mininet. Incoming attack packets were classified as either threat (1) or no threat (0). A dataset was created through tests on the existing network, and a mixed dataset was obtained by combining this dataset with another. ANN, Random Forest, XGBoost, AdaBoost, and GradientBoost techniques were used in classification processes, and performance comparisons were made between these techniques. It has been determined that the performance of the algorithms is mostly high and Random Forest has the highest accuracy rate on the dataset.

In their study, Peter B. et al. [26] research to ensure equal distribution of the load between proxy servers in the content distribution network (CDN). They developed an AI model utilizing the flexible structure of SDN and their ability to understand the general state of the network. The research involved obtaining bandwidth, number of hops, and other network parameters along the path, which were then used as input for the AI model. The data obtained was trained with Naive Bayes, Decision Tree, Random Forest, Logistic Regression, and K-nearest neighbor algorithms. In the training results, the Decision Tree algorithm is more successful than other algorithms and can make predictions with a 97% accuracy rate. According to the research results, the classical nearest server approach is compared with the AI-supported load-balancing system, and the positive effects of AI on the system are observed.

Thabo S. et al [27] present a comprehensive survey of load-balancing methods. According to their research, they examine load balancing under five main headings. These are controller load balancing, server load balancing, load balancing on wireless connections, communication path selection load balancing, and AI-based load balancing. In these subheadings, studies conducted in the general literature are discussed and the weaknesses and strengths of the studies are argued. In addition, at the end of the research, the performance criteria and application tools used in the projects were examined. In research, 93% of the studies use delay, 87% throughput, 100% QoS, and 73% complexity parameters. When the application tools are examined, 55% of the researchers test their proposed approaches in the Mininet environment. The most commonly used tool to generate network traffic is iperf3.

Cui C. and Xu Y. [28] aim to distribute the load equally among different subnets by taking advantage of the global view of SDN. In their study, an ANN is developed using bandwidth usage rate, packet loss rate, transmission delay, and hop count. The developed AI model is compared with the static round-robin algorithm. The results show that the network delay decreases by 19.3% at most and it is determined that there is an improvement in the system.

Şükran D. and Derya Y. [29], in their study utilizing the Floodlight controller as an SDN controller, processed the statistics obtained from packet monitoring on network traffic using MATLAB. The paper employs artificial intelligence optimization techniques such as linear search, tabu search, and simulated annealing. The analysis indicates that the traditional linear search method outperforms tabu search; however, the proposed hybrid algorithm yields more efficient results than linear search. It is observed that while AI techniques do not always guarantee the best performance, they offer significant improvements under suitable conditions. This study leverages artificial neural networks to predict packet flow in software-defined networks and conducts tests on different network topologies to identify the topology with the highest accuracy.

Anteneh A. Gebremariam et al. [30] focused on using machine learning technologies to solve complex problems in SDN and Network Functions Virtualization (NFV) based networks. The study examines the use of AI and ML in five main areas of SDN and NFV-based networks: network architecture, load balancing and resource utilization, fault detection and management, network management and operations, and network security. While the centralized structure of SDN and NFV offers flexibility in network management, it also introduces significant security risks if the central controller is compromised. The authors emphasize that machine learning can enhance network resilience in such environments by creating self-adaptive and self-managing systems. However, they also note that research in this field is limited and faces various challenges, highlighting the need for more in-depth studies to better understand and harness AI's potential in these networks. Table 1 lists the aims of the studies and the simulation programs reviewed in the literature. Some researchers did not perform simulations but only analyzed the data with different analysis methods.

Table 1. Summary of some relevant studies we reviewed.

| Author | Concept | Finding | Disadvantages |
| --- | --- | --- | --- |
| Warsama [20] | Creating a new bandwidth-based routing algorithm and comparing it with the classical ECMP algorithm. | The algorithm possesses global network knowledge and directs data flow in a non-congested direction. | Neglecting the CPU load and testing the algorithm on a small topology. |
| Bilal and Banu [21] | Research on load balancing in SDN-based data centers using artificial intelligence techniques. | Enabling dynamic load balancing with software-defined networks and enhancing system performance with artificial intelligence. | The scarcity of deep learning applications for network issues and insufficient data. |
| Ahmed and Ahmadreza [22] | A comprehensive investigation of AI-based load balancing in software-defined networks. | A comprehensive presentation of techniques and variables used since 2017. | In the examined studies, only a subset of QoS parameters are taken into account, while other criteria are overlooked. |

| Author | Topic | Contribution | Limitations |
|---|---|---|---|
| Mohamed et al. [23] | Server-based load balancing using software-defined networking. | Achieving better results with a bandwidth-based approach compared to classical load balancing algorithms. Testing the algorithm in both virtual and real environments. | Testing was conducted on very small topologies and showed a decline in performance after a certain threshold value. |
| Mosab et al. [12] | A comprehensive survey of load balancing techniques in software-defined networks. | Demonstrating that software-defined network-based load balancers are more successful than traditional load balancers and examining their differences. | The studies reviewed indicate security and scalability issues in the control plane. |
| Mohammad et al. [18] | Artificial Intelligence Based Reliable Load Balancing Framework in Software-Defined Networks. | Enhancing performance and enabling load balancing through modifications to existing artificial intelligence optimization techniques. | Algorithms are not being tested in large topologies and outside virtual environments. |
| Ekber [24] | Design of a controller based on an access protocol for eMBB traffic in Software-Defined Content Delivery Networks. | Development of a more successful routing algorithm based on the user access protocol in software-defined networks and explanation of the results using the PSNR metric. | Not considering CDN cache management and not comparing it with existing control algorithms. |
| Ayşe and Mevlüt [25] | Classification of security threats in software-defined networks using machine learning algorithms. | Enabling the automated detection of threats using artificial intelligence systems and demonstrating the high success of machine learning. | Insufficient explanation of the datasets used in training machine learning algorithms. |
| Peter et al. [26] | Classification-Based Load Balancing in Content Delivery Networks. | Developing an artificial intelligence model with a 97% success rate by obtaining bandwidth, hop count, and other network parameters along the path and observing positive effects in the system. | Insufficient explanation of the dataset used in training the models. |
| Thabo et al. [27] | Intelligent Load Balancing Techniques in Software Defined Networks: A Survey. | A comprehensive examination of load balancing in software-defined networks and identification of the most used network parameters and technologies. | It provides a general approach to artificial intelligence-based load-balancing techniques. It is a suitable article to gain information about algorithms and studies. Details can be learned by examining the mentioned studies. |
| Cui and Xu [28] | Development of an artificial intelligence-based load-balancing algorithm in software-defined networks. | Developing an artificial neural network using bandwidth utilization rate, packet loss rate, transmission delay, and hop count to improve the system. | Insufficient explanation of the dataset used in the training phase. |
| Sukran and Derya [29] | Software-Defined Networking Application with Artificial Intelligence Techniques. | Monitoring and optimizing network traffic using artificial intelligence techniques. | Artificial intelligence techniques do not always deliver the best performance. |

| Anteneh [30] | Applications of Artificial Intelligence and Machine Learning in the Area of SDN and NFV | A brief research presentation presents the main application areas of machine learning in SDN and NFV-based networks. | The limited research in the field and the potential for major security vulnerabilities due to possible errors. |
|---|---|---|---|

### 3. Machine learning-based routing approach in SDN-based architectures

SDNs are defined by a three-tier architecture consisting of the infrastructure, the control, and the application planes. There is a southbound interface between the infrastructure and control planes. The southbound interface most commonly uses the OpenFlow protocol. This protocol establishes communication between the switches and the controller. According to the protocol, when packet flows on the network arrive at the switch, it first checks its flow table. If the information about the packet is not in its flow table, it contacts the controller on the network and asks the controller to define a new rule. After the controller decides on the packet status, it updates the flow table of the relevant switch. The protocol is generally based on packet transmission according to the match-action status in the flow tables. There is a northbound interface between the control and application plane. The northbound interface provides communication between the programmer and the controller for programming. Therefore, the programmer can have an overview of the network, check the instant status of the devices in the network, and set new rules. Today, the lack of standardization in the northbound interface is one of the general problems of SDN [8].

Today, data analysis and machine learning play critical roles in many industries. When the complexity and volume of datasets increase, effective modeling, and accurate prediction-making become increasingly important. Thus, machine learning algorithms are used effectively, especially in classification tasks. In this study, the collected data are analyzed to identify the features that influence traffic levels. Subsequently, machine learning models are trained using these identified features. Training results and data analysis are shared in the research results section.

The routing algorithm is run before video transfer between the client and the server. In the first stage, the algorithm loads the AI model and prepares to make predictions. After the AI model is loaded into the system, the five shortest paths between the client and server are taken from the Floodlight controller. Each path is set with StaticEntryPusher, the necessary switches are added to the flow table, and the flow statistics between the 2 devices on the specified path are collected. The collected statistics are given as input to the AI model and a resulting traffic level is obtained. The resulting traffic levels are recorded in a dictionary with the route index. After the same process is performed for all routes, the route with the lowest traffic level is selected. By updating the flow table of the relevant switches on the selected route, video transfer between the client and server is ensured over that route. The pseudocode of the mentioned routing algorithm is given in Table 2. The NSFNET topology and scenario used in this study are shown in Figure 1.

Table 2. AI-based routing algorithm pseudocode

| Algorithm: | AI-based routing algorithm |
|---|---|
| Input: | Client and server IPv4 addresses |
| Output: | Setting the path with the lowest traffic level between the client and server |
| 1: | **function** *RoutingAlgorithm(client_ipv4, server_ipv4)* |
| 2: | $model \leftarrow LoadTheClassificationModel()$ |
| 3: | $paths \leftarrow GetTheShortestPaths(client\_ipv4, server\_ipv4)$ |
| 4: | $pathsWithTraffic \leftarrow$ Declare a new array |
| 5: | **for** *path* **in** *paths* **do** |
| 6: | $AddToFlowTable(path)$ |
| 7: | $pathStatistics \leftarrow GetThePathStatistics()$ |
| 8: | $prediction \leftarrow model.Prediction(pathStatistics)$ |
| 9: | $pathsWithTraffic.Append(prediction)$ |
| 10: | $DeleteAddedFlows(path)$ |
| 11: | **end for** |
| 12: | $path \leftarrow FindThePathWithLowestTraffic(pathsWithTraffic)$ |
| 13: | $AddToFlowTable(path)$ |
| 14: | **return** *path* |
| 15: | **end function** |

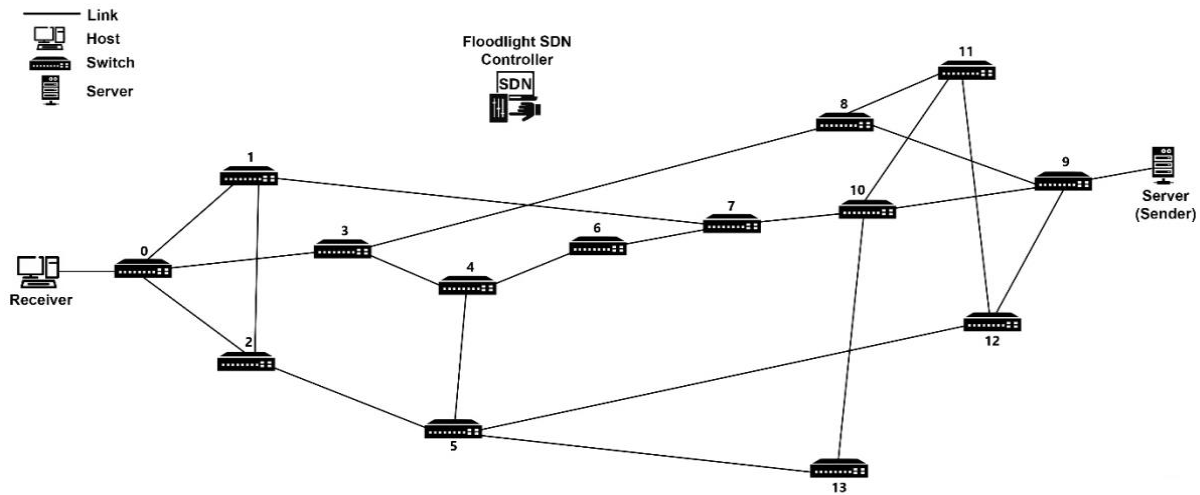

Figure 1. NSFNET topology created in the Mininet

This study's development environment utilized the Ubuntu 20.04 operating system on a machine with an Intel I7 quad-core processor and 8GB RAM, leveraging Floodlight as the SDN controller, Mininet for simulation, Iperf3, NFStream and Ping for traffic analysis, Google Colab for the machine learning platform, and Python3 as the programming language.

Mininet is an open-source emulation and testing platform for network protocols and applications. 55% of the studies in the literature on load-balancing algorithms in SDNs are done in the Mininet environment [27]. Mininet allows us to create virtual network topologies and simulate network protocols without physically using real network hardware. Since we will work in a virtual network environment without the need for physical devices, developing and testing reduce the cost. Some basic topology types to create in Mininet are Single Switch, Linear, Tree, and Full Mesh. Additionally, Python API allows the creation of the desired topology on Mininet. Any IP controller can be integrated into the created topology and managed with the help of the controller [31].

Floodlight, an open-source SDN controller, empowers network administrators with dynamic resource management and programmability. Leveraging the OpenFlow protocol, it controls network traffic, manages switches, and offers a REST API for developers to create flexible and scalable SDN applications in any programming language [32].

The National Science Foundation Network (NSFNET), which operated from 1985 to 1995, served as a vital research and education network within the United States. Its core network and interconnected access points (IPOPs) facilitated high-speed data transmission between universities, research institutions, and government agencies. Notably, NSFNET played a critical role in Internet development by utilizing and standardizing the TCP/IP protocol, laying the foundation for the broader Internet infrastructure that emerged in 1995 [33,34]. In this study, different scenarios are developed by leveraging the complex topology of NSFNET. The presence of alternative paths between different network devices and the variation in hop counts between these paths are key factors in designing the scenarios. NSFNET is chosen due to its sufficient network complexity and recognition in the literature.

To enhance the realism of the network environment, we continuously generate traffic between clients using iperf3. This versatile tool, commonly used for network performance testing, acts as a server-client model to simulate data transmission and measure key metrics like connection speed, latency, and packet loss. In our study, different clients are positioned on the path between the server and the client. TCP packet transfer is carried out between these clients. TCP packet transmission is carried out between these clients. The primary reason for utilizing TCP is its ability to provide reliable data transmission and ensure that data packets are correctly received by the destination. Through its error control and flow control mechanisms, TCP minimizes losses and errors during data transmission. This allows us to realistically evaluate the overall performance of the network and accurately simulate challenges such as potential delays and packet loss. Additionally, during TCP measurements conducted with the iperf3 tool, critical information such as the amount of data transferred and bandwidth for each transmission interval is obtained. These details are crucial for assessing network performance and enable a more detailed analysis of the network's data transmission capacity, latency, and efficiency. An example topology is shared in Figure 2.

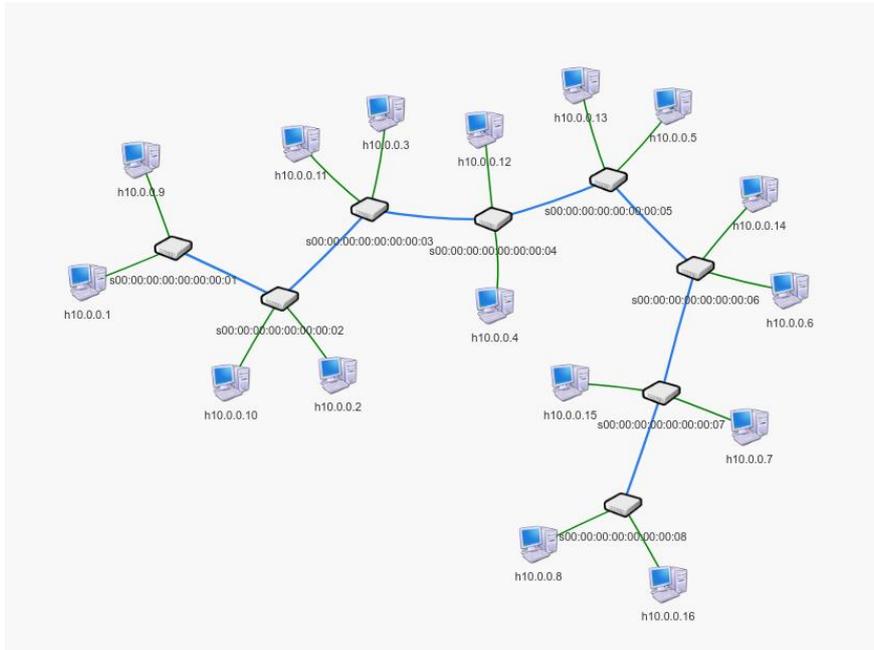

Figure 2. Example linear topology in Mininet environment

***Video transfer between Server and Client***

FFmpeg is an open-source multimedia framework. It provides tools and libraries for encoding, decoding, converting, and publishing audio, video, subtitles, and other multimedia files. FFmpeg supports many different multimedia formats and has a large user base. This framework contains a set of commands available via the command line. In the tests performed in our study, a video with .mp4 extension is converted to .ts (Transport Stream) extension with the "ffmpeg -i video.ts output.ts" command of the FFmpeg library. TS is a transport format that contains video and audio data. While the number of frames sent from the original video (video.ts) is 3990, the number of frames of the video received by the server (input.ts) is 3881. 9 frames were lost during transmission. The representation of normal and corrupted frames in the video obtained after the transmission is provided in Figure 3.

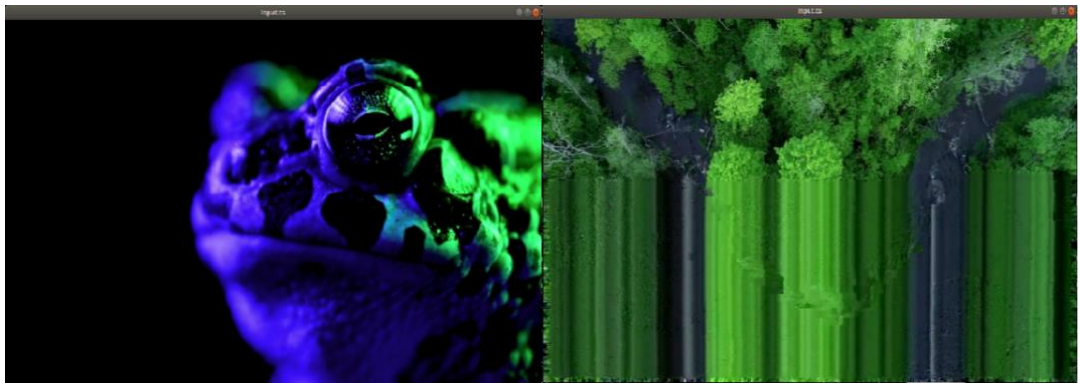

Figure 3. Normal and corrupted frames in the video obtained after the transfer

NFStream is a Python framework designed to make working with network data easy and intuitive [35]. It offers the opportunity to analyze data on the network practically. In our study, an interface on the network is listened to with NFStream, and statistical data about the packets is obtained. It creates a copy of existing packets while listening and does not interfere with the network. In the AI model, we have established, that 33 features are obtained with NFStream, and the effects of these features on the traffic level on the network are examined. In our study, when creating the NFSteamer object, the idle_timeout property was set to 15 seconds and statistical_analysis was set to True. Other features are used by default. The system flow chart developed while designing the proposed architecture is shown in Figure 4.

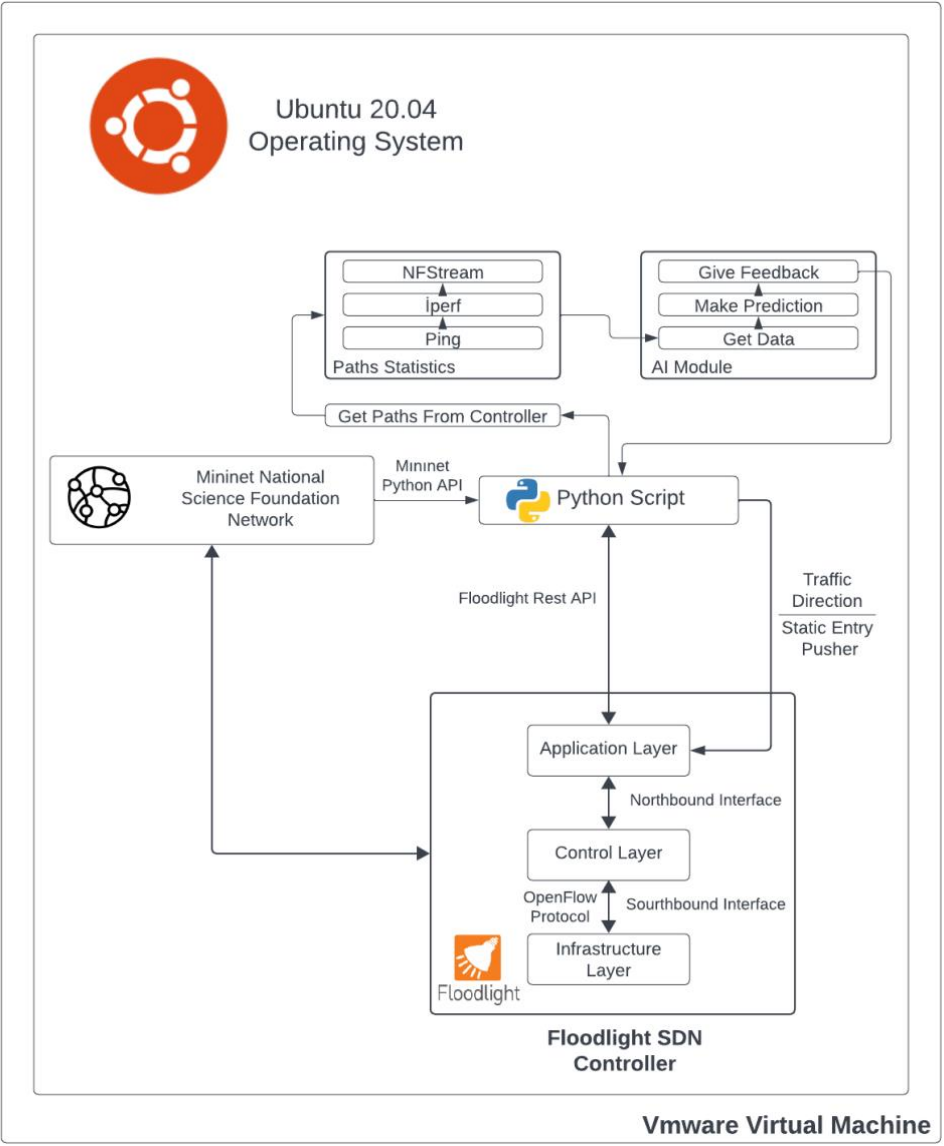

Figure 4. Proposed system flow diagram

**Creating Video Streaming Scenarios on the NSFNET Network**

Certain scenarios have been created to obtain data on the NSFNET network established in the Mininet environment. The scenarios are based on data transmission on 3 different paths on the network.-Each path contains a different number of hops, and this variation forms the foundation of the scenarios within the simulation environment. The scenarios are structured based on changes in hop count, established connections, and traffic levels. Two different traffic levels are created for each route and data transmission is made according to these traffic levels. The big picture of the network for scenarios is constructed and the devices are shown in Figure 1 in the introduction section. 14 switches, 1 client, and 1 server are in the network. There is also 1 Floodlight SDN Controller that controls all switches on the network. These network devices form the basis of the fictional scenarios. While designing scenarios, different clients are added to the network and traffic is created between them.

- *Video Transfer - First Scenario*

The first scenario is based on the shortest path between the client and the server. The shortest route on the mentioned network follows the switches 0 – 3 – 8 – 9 and reaches the destination with 3 hops. The first scenario is presented in Figure 5.

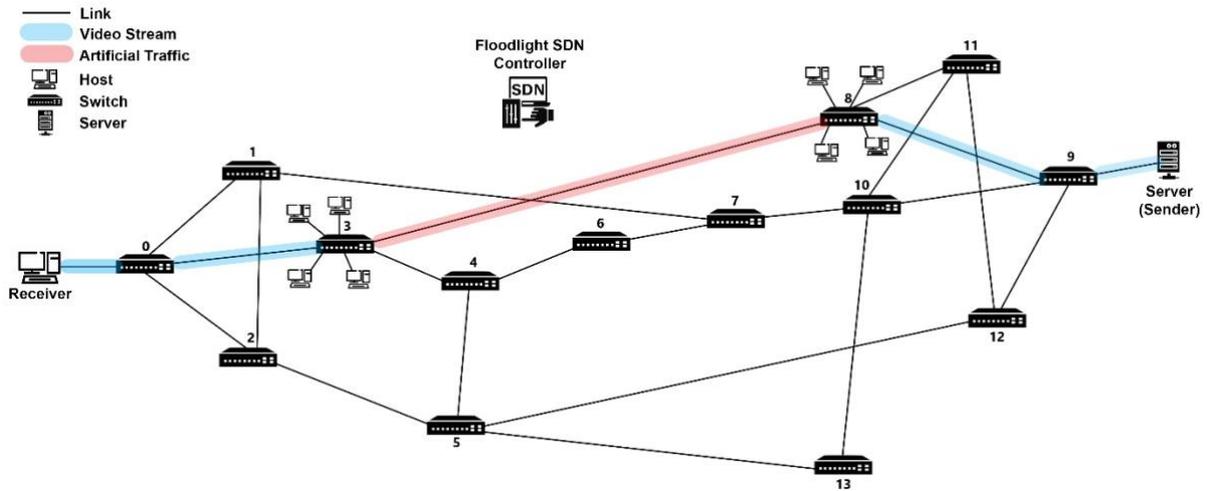

Figure 5. Visualization of the first video transmission scenario

In the scenario, 2 different traffic levels are created between the receiver and the sender, and 4 clients are assigned to switches 3 and 8. In a low-traffic environment, 4 selected clients establish 2 connections between each other, creating TCP traffic and using up to 30% of the bandwidth. In a high-traffic environment, 8 clients create TCP traffic by establishing 4 connections using approximately 90% of the bandwidth. Once traffic is generated, the video is transferred between the main client and the server. Details on client configurations for generating low and high traffic levels are given in Tables 3 and 4.

Table 1. Example traffic information in a low-traffic environment

| LOW TRAFFIC ENVIRONMENT | | | |
|---|---|---|---|
| SOURCE IP | DESTINATION IP | BANDWIDTH | PROTOCOL |
| 10.0.0.9 | 10.0.0.4 | 1500K | TCP |
| 10.0.0.17 | 10.0.0.15 | 1500K | TCP |

Table 2. Example traffic information in a high-traffic environment

| HIGH TRAFFIC ENVIRONMENT | | | |
|---|---|---|---|
| SOURCE IP | DESTINATION IP | BANDWIDTH | PROTOCOL |
| 10.0.0.9 | 10.0.0.4 | 2250K | TCP |
| 10.0.0.17 | 10.0.0.15 | 2250K | TCP |
| 10.0.0.18 | 10.0.0.16 | 2250K | TCP |

| 10.0.0.21 | 10.0.0.19 | 2250K | TCP |

- *Video Transfer - Second Scenario*

In the second scenario, the number of hops is increased by one and a path of 4 hop length is preferred. Video transfer between client and server follows the switches 0 – 1 – 7 – 10 – 9. Two different traffic levels are created on this path and the visualization of the scenario is shown in Figure 6.

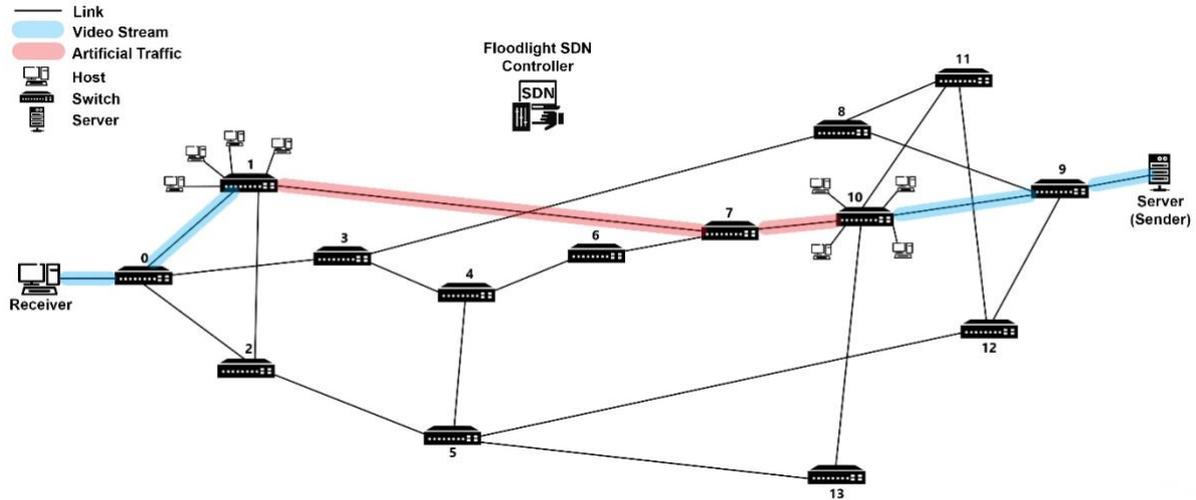

*Figure 6. Visualization of the second video streaming scenario*

In the second scenario, 4 clients are assigned to switches 1 and 10. In a low-traffic environment, it creates TPC traffic by establishing 2 connections between 4 clients, similar to the first scenario, and uses up to 30% of the bandwidth. In a high-traffic environment, 8 clients create TCP traffic by establishing 4 connections using approximately 90% of the bandwidth. In both cases, tests are performed by transmitting video.

- *Video Transfer - Third Scenario*

In the third scenario, the number of hops is increased to a path with 5 hops. Video transfer between the client and the server follows the switches 0 – 2 – 5 – 13 – 10 – 9. Similarly, low and high traffic is generated in this scenario using same number of clients and the same amount of bandwidth. Clients to generate traffic are assigned to switches 2 and 10. The scenario is shown in Figure 7.

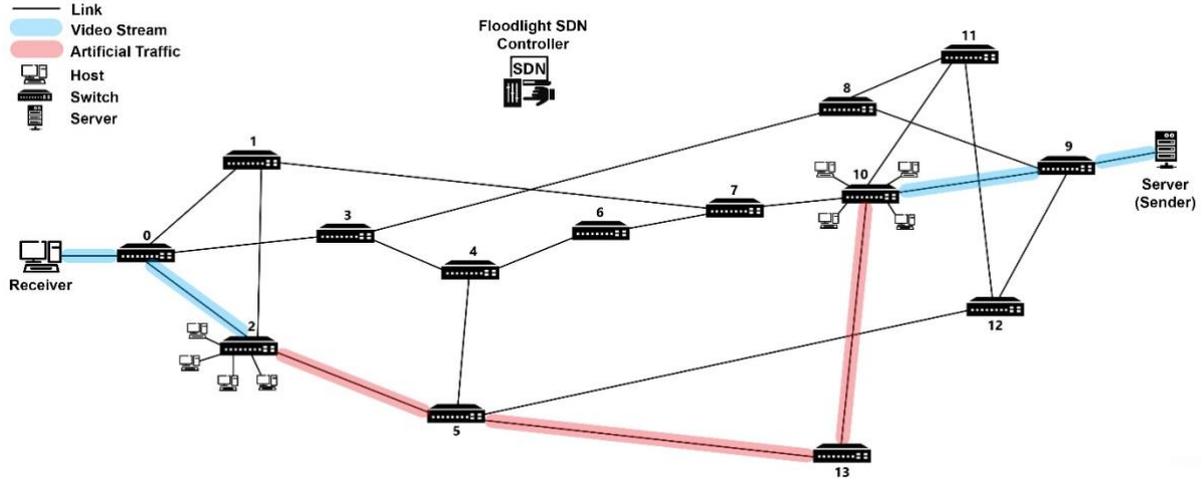

*Figure 7. Visualization of the third video transmission scenario*

4. **Measuring Video Quality after Video Transfer**

Video is transferred between the client and the server based on the scenarios. The transferred video is 66 seconds long and has a resolution of 1920x1080 pixels. The PSNR (Peak Signal-to-Noise Ratio) and SSIM (Structural Similarity Index) parameters are employed to measure the effects of the existing traffic on the transmission path and the video transmission protocol (TCP or UDP) on video quality. These metrics play a crucial role in evaluating the quality of the transmitted video compared to the original. PSNR and SSIM are considered key metrics for assessing transmission performance. To these metrics, the impacts of different traffic levels and protocols on video quality can be analyzed with concrete data [36].

PSNR is a commonly used metric to evaluate video quality. This metric measures how successful the processed version of an original video is against processes such as format changes, compression, or transmission over the network. To calculate PSNR, the first step is to calculate MSE (Mean Square Error) (E.1.1), the average of the squared differences of pixel values in the referenced original video and the processed video frames. Then, using this MSE value, video quality is measured with the PSNR (E.1.2) Equation.

$$\text{MSE} = \frac{1}{m \times n} \sum_{i=0}^{m-1} \sum_{j=0}^{n-1} [\, O(i,j) - D(i,j) \,]^2 \tag{E.1.1}$$

$$\text{PSNR} = 10 \cdot \log_{10}\left(\frac{MAX_1^2}{MSE}\right)$$

$$= 20 \cdot \log_{10}\left(\frac{MAX_1}{\sqrt{MSE}}\right)$$

$$= 20 \cdot \log_{10}\left(\frac{MAX_1}{\sqrt{MSE}}\right) \tag{E.1.2}$$

In this equation, MAX represents the peak value of pixels and is generally considered 255 when pixels are expressed as 8 bits. The equation expressing these calculations is (E.1.3):

$$\text{PSNR} = 48.1308036087 - 20 \cdot \log_{10}\left(\frac{MAX_1}{\sqrt{MSE}}\right) \tag{E.1.3}$$

SSIM is a metric that compares the similarity between two images. SSIM aims to obtain results closer to the human visual system and includes measurements such as structural, brightness, and contrast similarity between images.

It takes values between -1 and 1. Values approaching 1 indicate high similarity between two images, while values approaching -1 indicate low similarity. The equation expressing these calculations is (E.1.4):

$$SSIM(x, y) = \frac{(2\mu_x\mu_y + C_1)(2\sigma_{xy} + C_2)}{(\mu_x^2 + \mu_y^2 + C_1)(\sigma_x^2 + \sigma_y^2 + C_2)}. \tag{E.1.4}$$

• x and y are the two images being compared.

• μx and μy are the average values of the pixel values of the images.

• σx and σy are the standard deviations of the pixel values.

• σxy is the covariance between x and y.

• C1 and C2 are small constant values for stability and normalization.

Information about the transmission path between the client and the server is obtained using different tools (iperf3, NFStream, ping). A total of 49 features are collected regarding the transmission path using the aforementioned tools. However, not all of this features is directly related to the traffic level on the path. Through conducted analyses, the features influencing the traffic level have been identified and provided as input to the machine learning algorithm employed in this study. The results of these analyses reveal the effects related to traffic levels both before and after the transfer and are detailed comprehensively in the conclusion section of the study. The tools used to obtain this data, along with their names and descriptions, are presented in Table 7 (This table is in the Supplementary File).

**Generating Data with the Scenarios**

Data is produced with scenarios designed in the Mininet environment. A program was written in Python to generate data automatically. Mininet and other tools support Python because it's easy to use. The program flowchart is summarized in Figure 8. The network topology is created, Floodlight Controller configuration settings are made, and the paths for the scenario are set with Floodlight StaticEntryPusher. Then, traffic is created with iperf3, the parameters are obtained with the previously mentioned tools and video transfer is done and recorded between the client and server. Video is tested before and after transmission, and finally, the system is reset and run cyclically. Threads are created multiple times within the program to perform multiple operations.

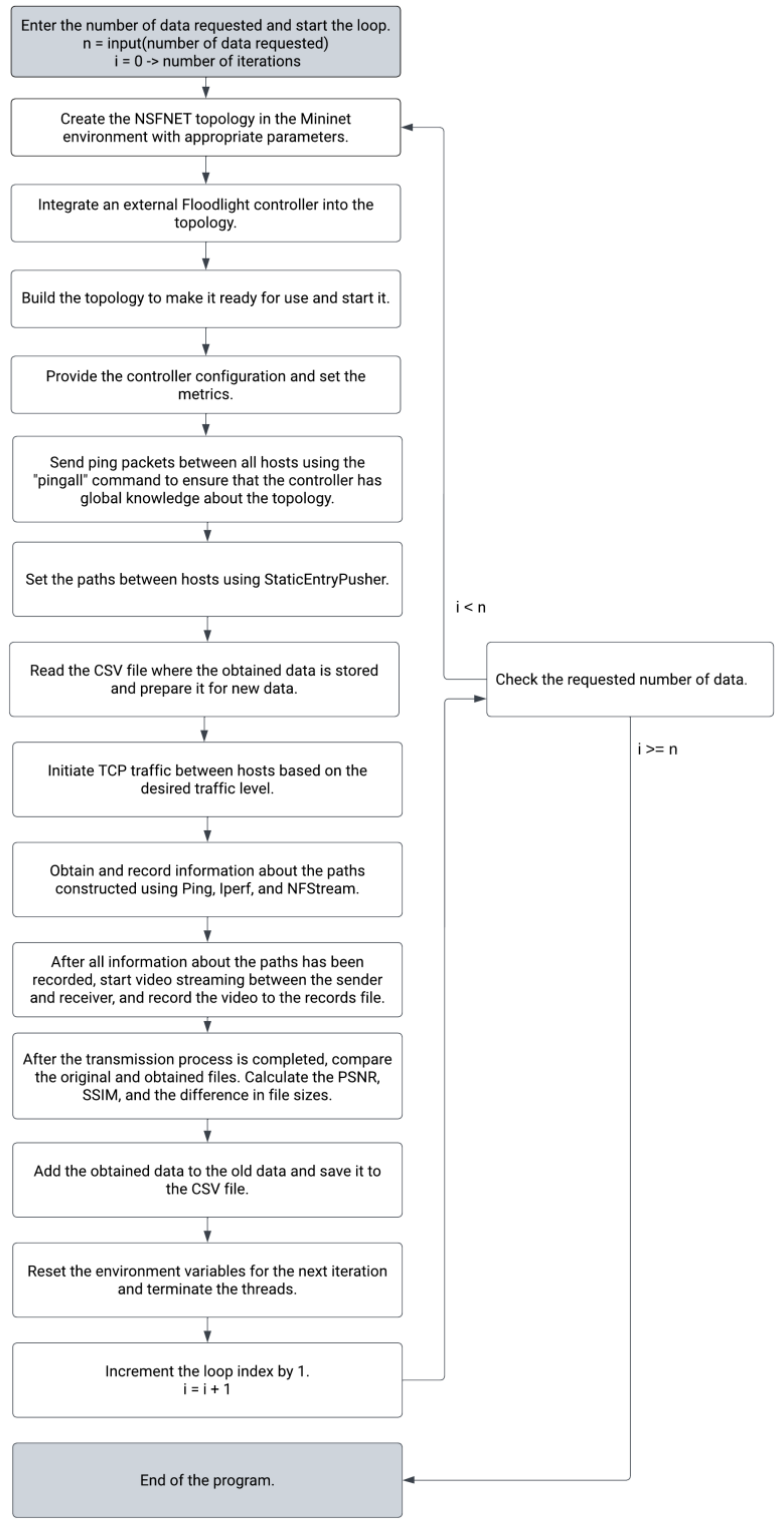

*Figure 8. Data generation process flowchart*

## 5. Performance Results

In the simulation environment, a total of 867 records are generated across three different scenarios. The data generation process is conducted in two phases for each scenario, representing low and high traffic conditions. Each record contains 50 features, with 49 representing various metrics collected during and before the video transmission, and last feature (label) indicating the traffic type in the environment. The data collection process is carried out using NFStream, iperf3, and ping tools to measure the connection quality between two devices in the network. These tools provide detailed insights into network performance by collecting bi-directional traffic statistics such as average round-trip time (RTT), packet loss, bits per second (throughput), bandwidth utilization rate, and other relevant metrics. The distribution of data based on traffic levels and the number of hops in each scenario is illustrated in Figure 9.

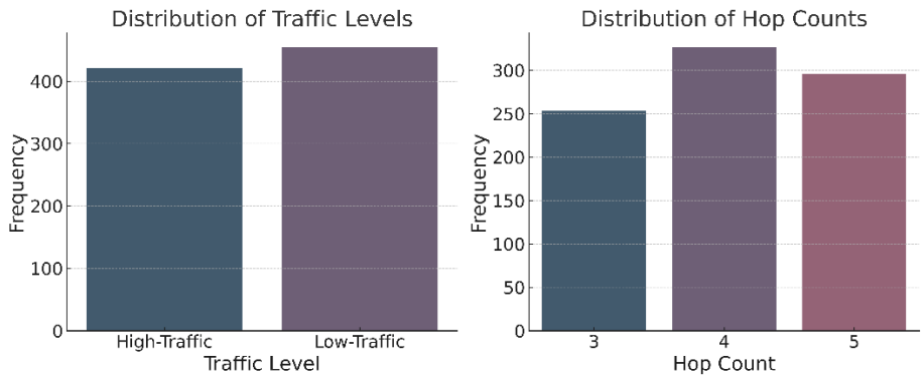

Figure 9. Distribution of data based on traffic level and hop count.

The size of the video transferred in high and low-traffic environments varies and data loss occurs. The original video size used during data generation is 39478308 bits. When the video is transferred in a low-traffic environment, the average transferred file size is 39320447.71 bits, and 157860.28 bits are lost. It is observed that there is approximately 0.4% data loss. When the video is transferred in a high-traffic environment, the average transferred file size is 36388202.41 bits, and 3090105.58 losses occur. It is observed that there is approximately 7.82% data loss.

Upon analyzing the effects of traffic type on different features, it is observed that traffic type has a direct impact on 19 features. During the analysis process, the correlation values between these features and the traffic type are examined. Based on the obtained correlation values, the features with the highest correlation to traffic type are identified as average RTT, packet loss, bandwidth utilization rate, and bits per second (throughput), respectively. In line with these analysis results, the confusion matrix illustrating the relationship between traffic type and the 19 features is presented in Figure 10.

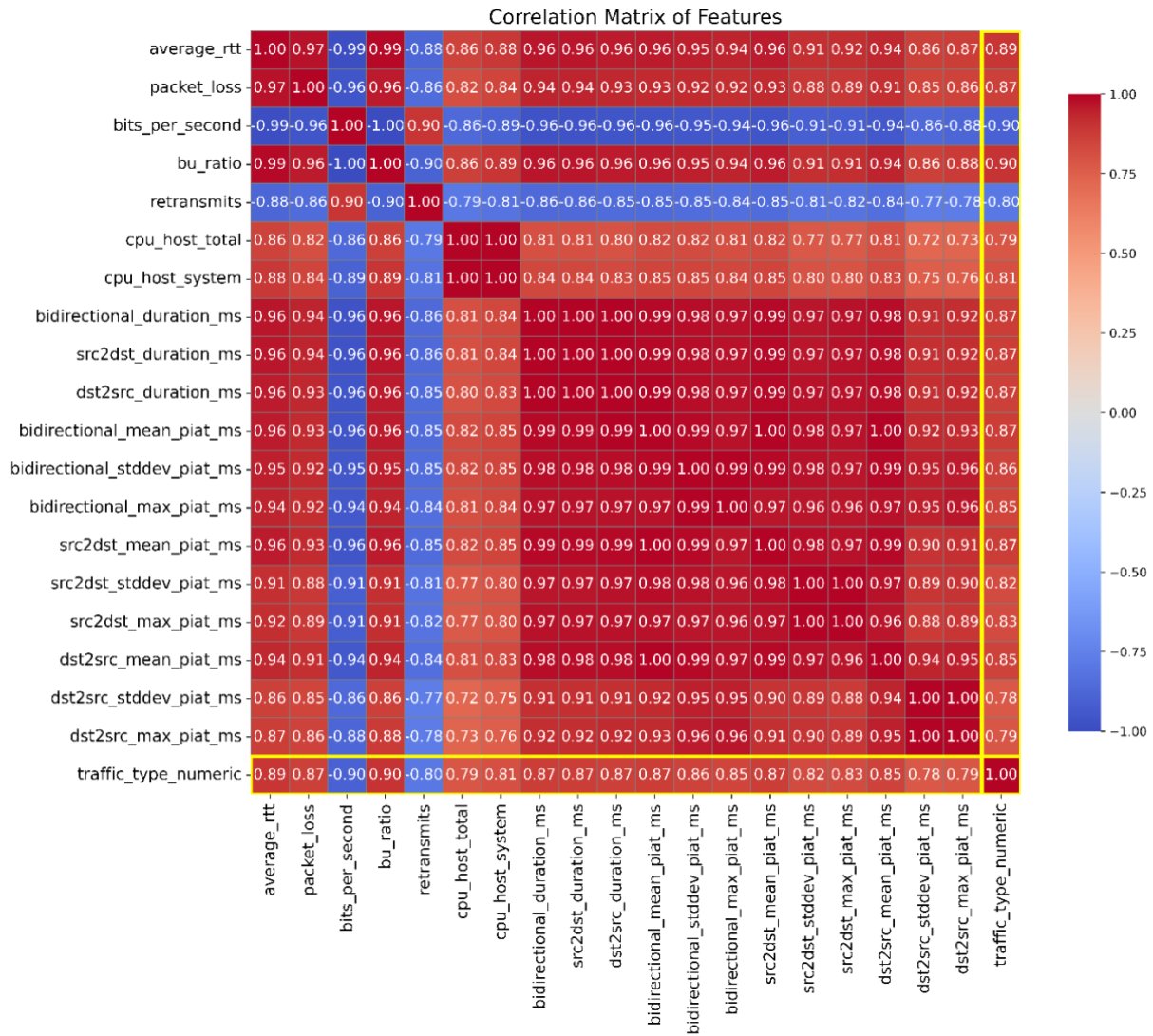

Figure 10. Confusion matrix illustrating the correlations among the 19 different features in the dataset.

To further examine the distribution of the four features that show the highest correlation with traffic type and their relationships with traffic conditions, box plots for these features are presented in Figure 11. The results indicate that the average RTT in low-traffic conditions is 306.27 ms, while this value rises to 1710.59 ms in high-traffic conditions, demonstrating a significant increase in network latency under heavy traffic.

Similarly, packet loss is observed as 0% in low-traffic environments, whereas it reaches 13.99% under high-traffic conditions, indicating a substantial increase in transmission losses with higher traffic levels. The bandwidth utilization ratio is detected to be 33.18% on average in low-traffic environments, it increases to 93.31% under high-traffic conditions, revealing that the network's bandwidth becomes heavily saturated under these circumstances.

Lastly, the number of bits transmitted per second averages 6.68 Mbps in low-traffic environments, while this value drops to 0.67 Mbps in high-traffic conditions. This finding illustrates a significant reduction in data transmission speed under heavy traffic. The box plots presented in Figure 11 visually depict the impact of traffic levels on specific features and allow for a detailed examination of how the data varies under different traffic conditions.

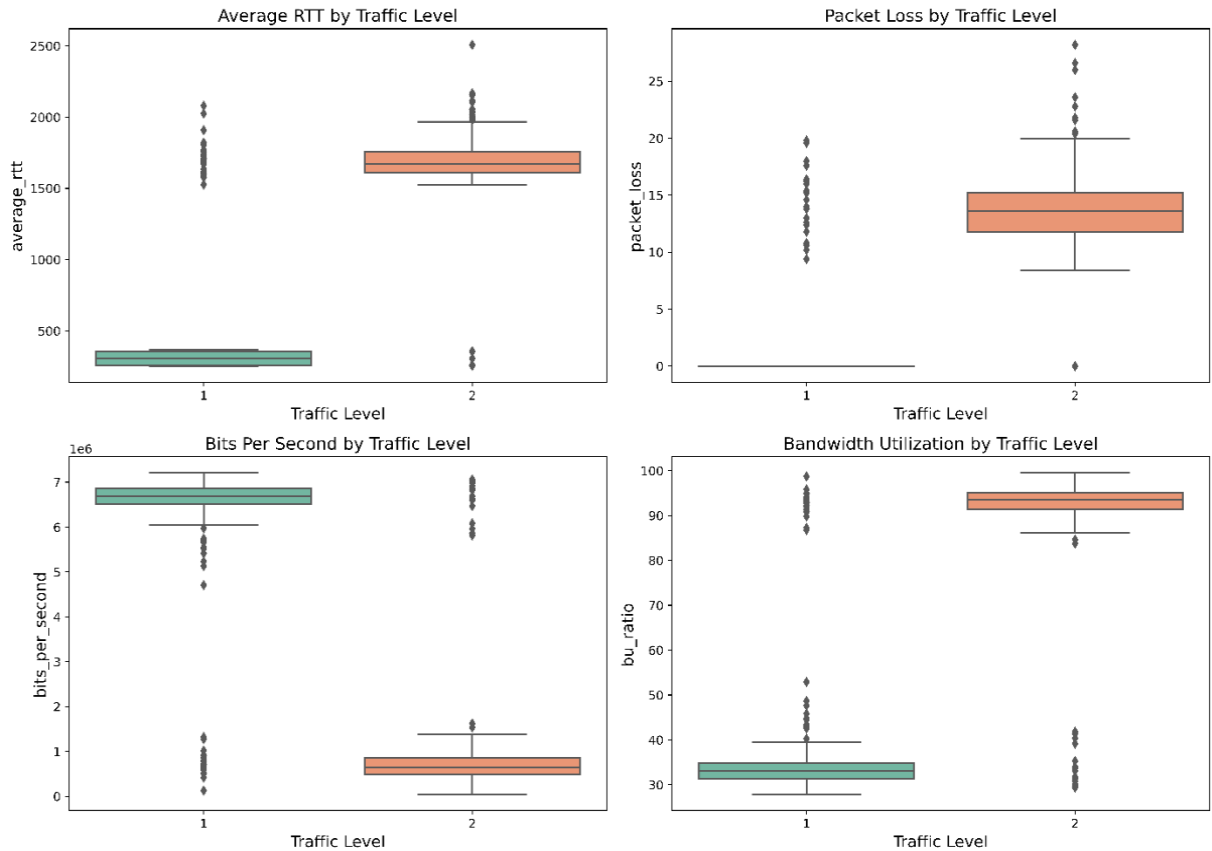

Figure 11. Box plots show the relationship between traffic type and the features of average RTT, packet loss, bandwidth utilization rate, and bits per second.

PSNR and SSIM parameters are used to measure the effect of the TCP traffic constructed in the scenario on the transmitted video quality. These parameters give us information about how different the video is from the original video. High values indicate that the video quality is close to the original, while low values indicate distortion in the video frames. Upon examining the data, the average PSNR value for videos transmitted in low-traffic environments is calculated to be 36.45 dB. This result indicates that the video quality is quite close to the original. In contrast, the average PSNR value drops to 21.97 dB in high-traffic environments, signifying a noticeable decline in video quality. In terms of SSIM values, the average SSIM for videos transmitted in low-traffic environments is initially measured at 0.996, indicating that the video quality is highly similar to the original. However, in high-traffic environments, this value decreases to 0.917, indicating a significant reduction in the structural similarity of the video. These results reveal that traffic levels have a significant impact on video quality, with higher quality video transmission occurring under low-traffic conditions, whereas quality degradation increases under high-traffic conditions. In the collected data, the average PSNR and SSIM values of the transmitted videos in low and high-traffic environments are shown in Figure 12.

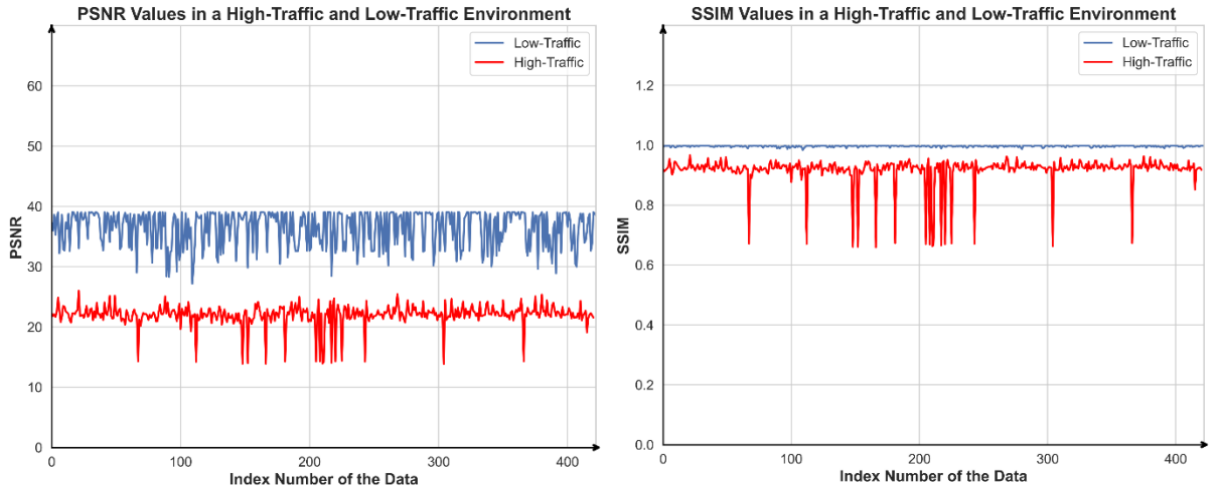

*Figure 12. PSNR and SSIM values based on traffic level.*

Data is collected using NFStream, iperf3, and ping tools to measure the connection quality between two devices in the dataset. When the collected data is examined, it is observed that all features do not directly affect the result value, and features that do not have an effect are removed from the dataset. A total of 33 features were obtained with the NFStream tool. The traffic level of these features was examined by creating a correlation matrix and it was observed that 12 features affected the result feature. Features that have no effect are generally related to packet sizes. When 12 features collected with other tools are examined, it is observed that 7 features affect the traffic level. As a result, the number of features, which was initially 49, was reduced to 19 by feature extraction as a result of data analysis. These 19 features constitute the key inputs for training the binary classification model. These selected features are carefully chosen to enhance the model's accuracy and reduce noise that could be introduced by irrelevant data. Thus, it is aimed to optimize the model performance and provide faster and more effective classification.

### 5.1 Binary Classification Training Results

After the analysis, the dataset is trained using various binary classification models. During the training process, the dataset is split into 70% for training, 15% for validation, and 15% for testing. Experiments are conducted using XGBoost, Random Forest, SVM, KNN, AdaBoost, and Logistic Regression models. According to the results, the overall performance of the models is quite high due to the dataset's structure and the clear distinction between features. The outcomes based on f1 score, precision, and recall metrics are presented in Table 5. These results demonstrate the overall performance of the models and their effects on the dataset.

Logistic Regression demonstrates the best results in terms of the F1 score. With an F1 score of 96.4%, Logistic Regression outperforms other models by exhibiting high performance in both precision and recall metrics. This result indicates that the Logistic Regression model has a strong ability to accurately predict the positive class (precision) and effectively classify without missing true positives (recall).

Table 5. Performance results of machine learning models

| Model | Precision | Recall | F1 Score |
|---|---|---|---|
| XGBoost | 0.956 | 0.956 | 0.956 |
| **Random Forest** | 0.943 | **0.971** | 0.957 |
| SVM | 0.955 | 0.941 | 0.948 |
| **KNN** | 0.917 | **0.971** | 0.943 |
| AdaBoost | 0.929 | 0.956 | 0.942 |
| **Logistic Regression** | **0.957** | **0.971** | **0.964** |

The success of the Logistic Regression model is primarily due to the compatibility of the dataset's structure with this model. The linear separability of features within the dataset enables the Logistic Regression model to effectively learn these features and utilize them efficiently during the classification process. This phenomenon is clearly observed in the confusion matrix of the Logistic Regression model. The confusion matrix presented in Table 6 provides a detailed insight into the classification performance of the model.

Table 6. Confusion matrix of logistic regression model

|  |  | Predicted Class | |
|---|---|---|---|
|  |  | Low Traffic | High Traffic |
| **Actual Class** | Low Traffic | 61 | 3 |
|  | High Traffic | 2 | 66 |

The ROC curve of the Logistic Regression model is a crucial visualization that evaluates the model's ability to distinguish between positive and negative classes. The ROC curve presented in Figure 13 confirms the overall performance of the model, resulting in a high AUC value of 0.955. This indicates that the Logistic Regression model is highly effective in distinguishing the positive class.

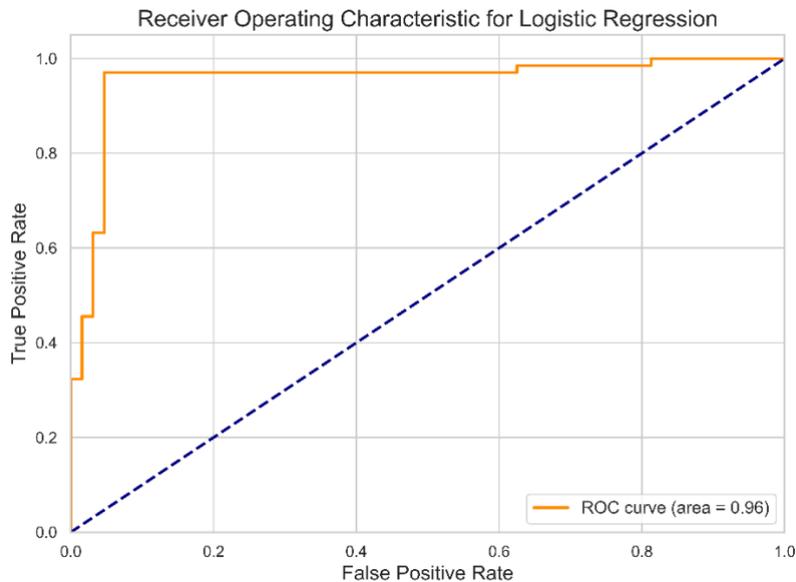

Figure 13. ROC (Receiver Operating Characteristic) curve for Logistic Regression model.

In conclusion, the Logistic Regression model demonstrates the highest performance among the binary classification models used in this study. Although other models also performed well, the Logistic Regression model maximizes overall classification success, particularly by offering a balanced precision and recall ratio. The structure of the dataset and the low noise level significantly contribute to this success. Synthetic data generation can involve less noise compared to real-world data, which generally enhances model performance. Moreover, during synthetic data generation, it is possible to create meaningful and distinct separations within the dataset, allowing linear models like Logistic Regression to learn these separations more effectively.

### 5.2 Performance of the Proposed Routing Algorithm

The proposed AI-based routing algorithm is tested in a scenario based on NSFNET topology. In the scenario, there are 19 clients generating traffic, 1 main client (receiving the video), and 1 server (sending the video). In addition, the Floodlight controller, which we describe as the brain of the network, is set to the hop count value as a metric. Figure 14 shows the traffic flows and degrees between hosts and provides a general visualization of the scenario.

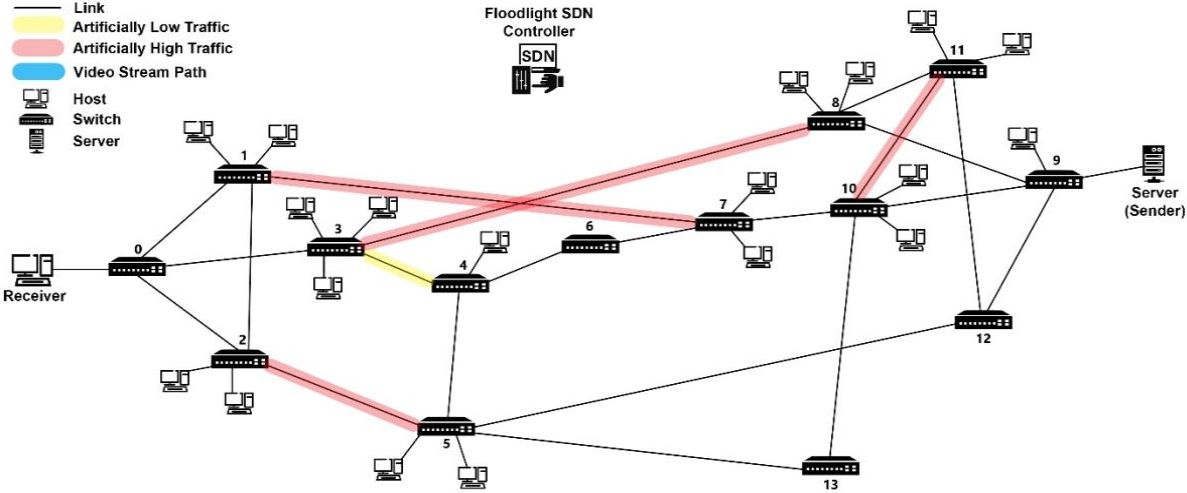

*Figure 14. Visualization of the scenario designed to test the performance of the proposed algorithm.*

During the testing phase, data is transferred between the client and the server without any intervention in the network first. The Floodlight controller performs data flow over the shortest distance with the hop count metric and follows the switches 0-3-8-9. Heavy traffic on this path negatively affects data flow and data losses occur. In previous tests, quality values were low when the video was sent on the path the controller had traveled. Figure 15 shows the path followed by the Floodlight controller during data transfer.

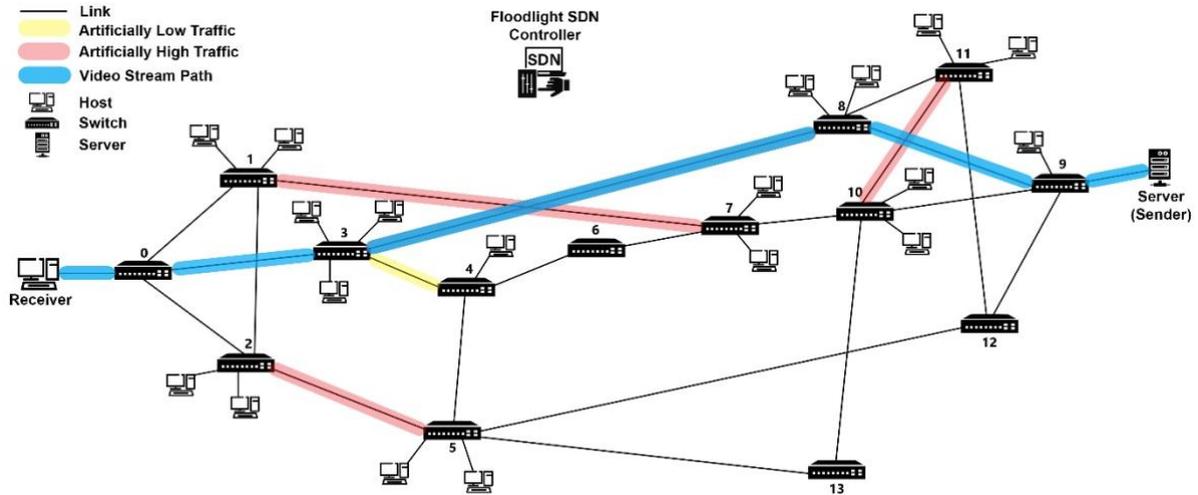

*Figure 15. Visualization of the path followed by the Floodlight controller based on hop count metric during data transfer.*

In the second part of the testing phase, the proposed routing algorithm is tested. The same environment variables as in the first stage are used for testing. Before the transfer between the client and the server, the proposed algorithm works and finds the route with a low traffic level. During the transfer, it follows the switches 0-3-4-5-12-9. The lower traffic on this path allows for more successful data transfer. Figure 16 shows the path that the proposed algorithm follows during data transfer.

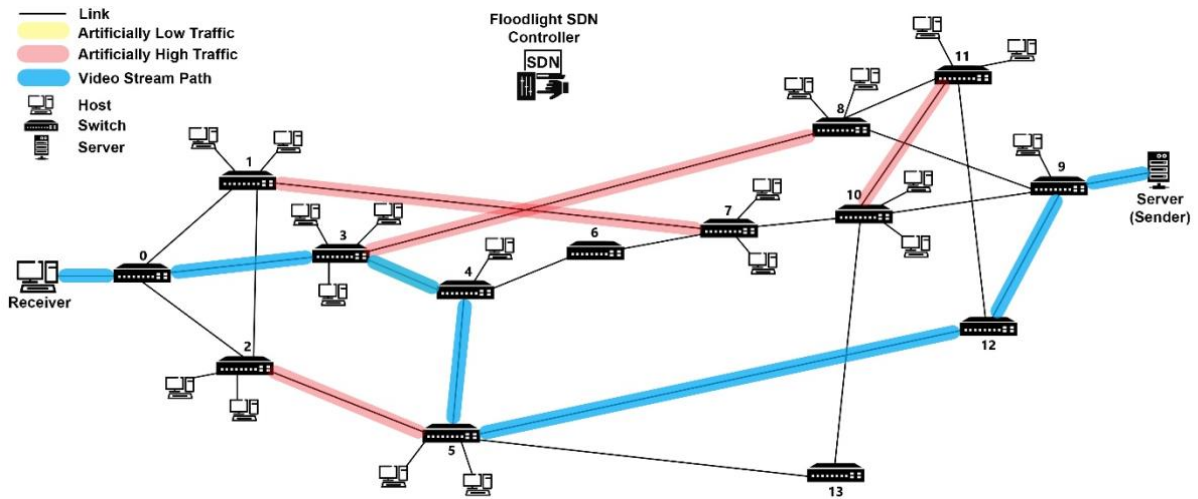

*Figure 16. Visualization of the path followed by the proposed algorithm during data transfer.*

After executing the scenarios of both algorithms, the data collected during the test are compared. For comparison, the network is listened to for 60 seconds and RTT and throughput values are recorded on the path. A comparison of the recorded data in Figure 17 is shown in line graphs.

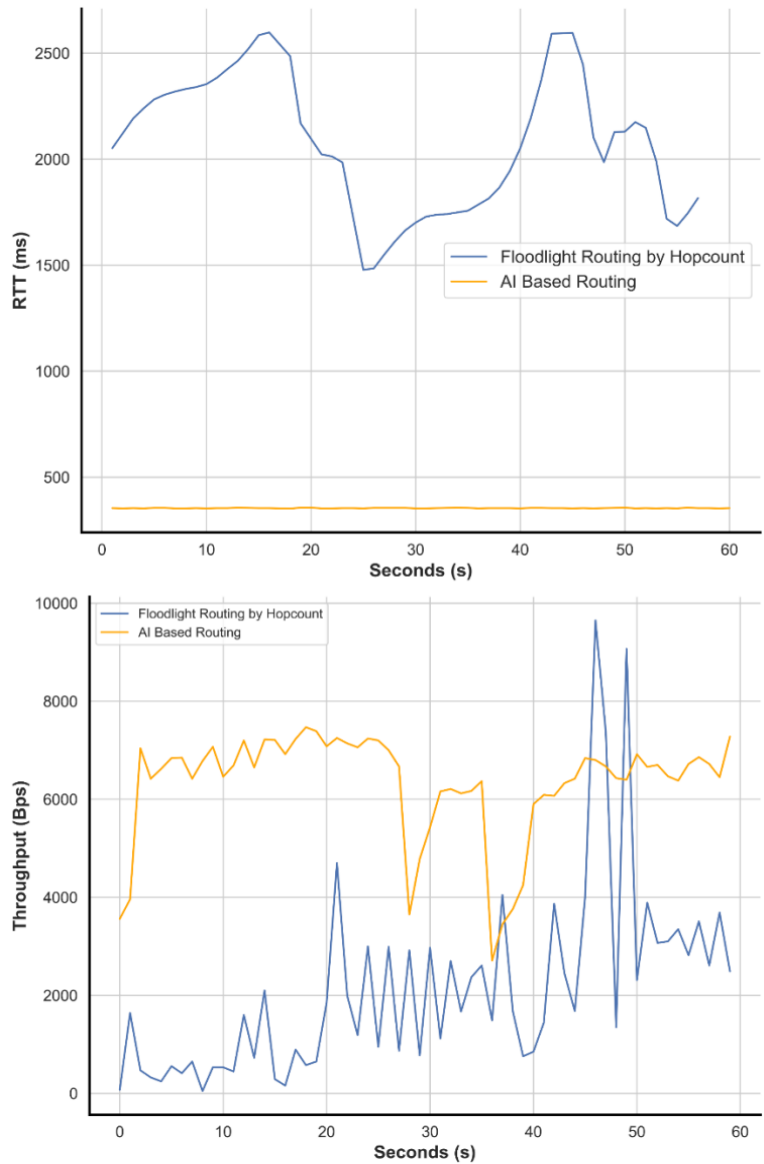

*Figure 17. Comparison of RTT and Throughput values between Floodlight hop count metric routing algorithm and the proposed algorithm.*

In the first test phase, as a result of the default routing algorithm with hop count metric, an average RTT of 2063.14 ms and throughput of 2135.90 bps are obtained. In the second stage, with the proposed algorithm, these rates are improved and 354.05 ms RTT and 6313.50 bps throughput values are obtained. As a result of the comparison, the algorithm we propose successfully detects the path with low traffic levels and transfers data through this path.

These findings demonstrate that the proposed algorithm significantly reduces RTT and meaningfully increases throughput. Additionally, the results indicate that the proposed algorithm successfully identifies paths with low traffic levels in the network, enabling more efficient data transmission through these paths. Notably, the marked reduction in RTT suggests that network latency is minimized, allowing data packets to be transmitted more quickly. Similarly, the substantial increase in throughput confirms that the network's data transmission capacity is optimized, leading to higher operational efficiency.

A more comprehensive evaluation of the results is also conducted in terms of video quality. Data transfers using the default algorithm yielded PSNR and SSIM values of 21.45 dB and 0.927, respectively, whereas the proposed AI-based routing algorithm increased these values to 37.27 dB PSNR and 0.997 SSIM. These results demonstrate that the proposed algorithm significantly enhances not only network performance but also user experience. The notable improvements in PSNR and SSIM values confirm that the quality of the transmitted video is substantially improved, offering a more satisfactory experience for users. In conclusion, the proposed AI-based routing algorithm outperforms the default algorithm in reducing latency and increasing data transfer rates and video quality.

## 6. Conclusion

The study focused on the advantages of SDN over traditional networks. In particular, a detailed examination of load balancers and routing algorithms in the data plane highlights the importance of advances in this field. SDN appears to be more successful and increases performance than traditional approaches. In the study, video transfer has been done between the selected client and server on the NSFNET topology. As a result of the traffic generated on the network during the transfer, some paths negatively affect the video quality due to traffic. By realizing the scenarios, the proposed AI-based routing algorithm is more successful than the controller hop count metric routing algorithm.

The study is conducted in the Mininet virtual simulation environment and has not been tested on a real network. While the virtual simulation environment has many advantages such as cost, time, and convenience for the testing environment, it has some disadvantages. The dataset created for the AI-based routing algorithm is completely artificial and different from real-time traffic because it is produced based on the scenarios in this study. For this study to be more efficient, testing it at real distances and in a real network environment will allow more accurate results to be obtained. During the design phase of the proposed algorithm, many different parameters are used to detect traffic on the network. These parameters help us detect the traffic situation on the network more clearly and from various dimensions. The main purpose is to monitor traffic levels across various metrics and more accurately detect both normal and anomalous conditions. At this stage, the computational capacity of AI and existing computers plays an important role. As a result, the findings can guide future studies on routing and AI optimization in SDN. With the innovative approach of SDN, programmers can have a say in the network and can easily perform operations and development.

**Acknowledgements**

This work was funded by the Scientific Research Unit of Bursa Uludag University (Project number FGA-2023-1557).

**References**

[1]     Cisco, Cisco Annual Internet Report (2018-2023), 2020.

[2]     D. Kreutz, F.M.V. Ramos, P.E. Verissimo, C.E. Rothenberg, S. Azodolmolky, S. Uhlig, Software-defined networking: A comprehensive survey, Proceedings of the IEEE 103 (2015) 14–76. https://doi.org/10.1109/JPROC.2014.2371999.

[3]     D. Kreutz, F.M.V. Ramos, P. Verissimo, Towards secure and dependable software-defined networks, in: Proceedings of the Second ACM SIGCOMM Workshop on Hot Topics in Software Defined Networking, 2013: pp. 55–60. https://doi.org/10.1145/2491185.2491199.

[4]     M. Cicioğlu, A. Çalhan, Energy-efficient and SDN-enabled routing algorithm for wireless body area networks, Comput Commun 160 (2020) 228–239. https://doi.org/10.1016/j.comcom.2020.06.003.

[5]     N. Feamster, J. Rexford, E. Zegura, The Road to SDN: An Intellectual History of Programmable Networks, ACM Sigcomm Computer Communication 44 (2014) 87–98. https://doi.org/10.1145/2602204.2602219.

[6]     N. McKeown, T. Anderson, H. Balakrishnan, G. Parulkar, L. Peterson, J. Rexford, S. Shenker, J. Turner, OpenFlow: enabling innovation in campus networks, ACM SIGCOMM Computer Communication Review 38 (2008) 69. https://doi.org/10.1145/1355734.1355746.

[7]     I.F. Akyildiz, A. Lee, P. Wang, M. Luo, W. Chou, A roadmap for traffic engineering in SDN-OpenFlow networks, Computer Networks 71 (2014) 1–30. https://doi.org/10.1016/j.comnet.2014.06.002.

[8]     F. Hu, Q. Hao, K. Bao, A Survey on Software Defined Networking (SDN) and OpenFlow: From Concept to Implementation, IEEE Communications Surveys & Tutorials (2014). https://doi.org/10.1109/COMST.2014.2326417.

[9]     B.A.A. Nunes, M. Mendonca, X.N. Nguyen, K. Obraczka, T. Turletti, A survey of software-defined networking: Past, present, and future of programmable networks, IEEE Communications Surveys and Tutorials 16 (2014) 1617–1634. https://doi.org/10.1109/SURV.2014.012214.00180.

[10]    M. Cicioğlu, A. Çalhan, MLaR: machine-learning-assisted centralized link-state routing in software-defined-based wireless networks, Neural Comput Appl 35 (2023) 5409–5420. https://doi.org/10.1007/s00521-022-07993-w.

[11]    M. Cicioglu, A. Calhan, A Multi-Protocol Controller Deployment in SDN-based IoMT Architecture, IEEE Internet Things J (2022) 1–1. https://doi.org/10.1109/JIOT.2022.3175669.

[12]    M. Hamdan, E. Hassan, A. Abdelaziz, A. Elhigazi, B. Mohammed, S. Khan, A. V. Vasilakos, M.N. Marsono, A comprehensive survey of load balancing techniques in software-defined network, Journal of Network and Computer Applications 174 (2021). https://doi.org/10.1016/j.jnca.2020.102856.

[13]    A. Drescher, A Survey of Software-Defined Wireless Networks, Washiongton University in St.Louis (2014) 1–15.

[14]    C. Yu, C. Lumezanu, A. Sharma, Q. Xu, G. Jiang, H. V. Madhyastha, Software-Defined Latency Monitoring in Data Center Networks, in: 2015: pp. 360–372. https://doi.org/10.1007/978-3-319-15509-8_27.

[15] A.-C. Anadiotis, L. Galluccio, S. Milardo, G. Morabito, S. Palazzo, SD-WISE: A Software-Defined WIreless SEnsor network, Computer Networks 159 (2019) 84–95. https://doi.org/10.1016/j.comnet.2019.04.029.

[16] E. Alpaydin, Introduction to Machine Learning Ethem Alpaydin., 2014.

[17] S.A. Latif, F.B.X. Wen, C. Iwendi, L.F. Wang, S.M. Mohsin, Z. Han, S.S. Band, AI-empowered, blockchain and SDN integrated security architecture for IoT network of cyber physical systems, Comput Commun 181 (2022) 274–283. https://doi.org/10.1016/j.comcom.2021.09.029.

[18] M.R. Belgaum, F. Ali, Z. Alansari, S. Musa, M.M. Alam, M.S. Mazliham, Artificial intelligence based reliable load balancing framework in software-defined networks, Computers, Materials and Continua 70 (2021). https://doi.org/10.32604/cmc.2022.018211.

[19] M. Latah, L. Toker, Application of artificial intelligence to software defined networking: A survey, Indian J Sci Technol 9 (2016). https://doi.org/10.17485/ijst/2016/v9i44/89812.

[20] A. Warsama, Traffic Engineering with SDN Optimising traffic Load-Balancing with OpenFlow, Digitala Vetenskapliga Arkivet (2020). https://www.diva-portal.org/smash/record.jsf?pid=diva2%3A1449012&dswid=9575 (accessed March 11, 2024).

[21] B. Babayigit, B. Ulu, Deep learning for load balancing of SDN-based data center networks, International Journal of Communication Systems 34 (2021). https://doi.org/10.1002/dac.4760.

[22] A.H. Alhilali, A. Montazerolghaem, Artificial intelligence based load balancing in SDN: A comprehensive survey, Internet of Things (Netherlands) 22 (2023). https://doi.org/10.1016/j.iot.2023.100814.

[23] M.I. Hamed, B.M. ElHalawany, M.M. Fouda, A.S.T. Eldien, A new approach for server-based load balancing using software-defined networking, in: 2017 IEEE 8th International Conference on Intelligent Computing and Information Systems, ICICIS 2017, 2017. https://doi.org/10.1109/INTELCIS.2017.8260023.

[24] E.Ç. Kük, Access protocol based controller desing for eMBB traffic in software-defined content delivery network, Master Science, Applied and Natural Sciences, 2021.

[25] A. Nur Temurçin, M. Ersoy, Investigation of AI-based network traffic management mechanisms in network infrastructure, International Journal of Sustainable Engineering and Technology 7 (2023) 31–40.

[26] P. Bakonyi, T. Boros, I. Kotuliak, Classification Based Load Balancing in Content Delivery Networks, in: 2020 43rd International Conference on Telecommunications and Signal Processing, TSP 2020, 2020. https://doi.org/10.1109/TSP49548.2020.9163470.

[27] T. Semong, T. Maupong, S. Anokye, K. Kehulakae, S. Dimakatso, G. Boipelo, S. Sarefo, Intelligent load balancing techniques in software defined networks: A survey, Electronics (Switzerland) 9 (2020). https://doi.org/10.3390/electronics9071091.

[28] C.X. Cui, Y. Bin Xu, Research on load balance method in SDN, International Journal of Grid and Distributed Computing 9 (2016). https://doi.org/10.14257/ijgdc.2016.9.1.03.

[29] S. Degirmenci, D. Yiltas, Software defined network application with artificial intelligence techniques, Journal of Engineering Sciences and Design 8 (2020) 999–1009. https://doi.org/10.21923/jesd.676110.

[30] A.A. Gebremariam, M. Usman, M. Qaraqe, Applications of Artificial Intelligence and Machine Learning in the Area of SDN and NFV: A Survey, in: 16th International Multi-Conference on Systems, Signals and Devices, SSD 2019, 2019. https://doi.org/10.1109/SSD.2019.8893244.

[31] Mininet, Mininet: An Instant Virtual Network on your Laptop (or other PC) - Mininet, Mininet.Org (2014).

[32] Floodlight, Floodlight, Floodlight Documentation (2016).

[33] J.D. Rogers, Internetworking and the politics of science: NSFNET in internet history, Information Society 14 (1998). https://doi.org/10.1080/019722498128836.

[34] Y.R. Chen, A. Rezapour, W.G. Tzeng, S.C. Tsai, RL-Routing: An SDN Routing Algorithm Based on Deep Reinforcement Learning, IEEE Trans Netw Sci Eng 7 (2020). https://doi.org/10.1109/TNSE.2020.3017751.

[35] Z. Aouini, A. Pekar, NFStream: A flexible network data analysis framework, Computer Networks 204 (2022). https://doi.org/10.1016/j.comnet.2021.108719.

[36] Sara, U., Akter, M., & Uddin, M. S. (2019). Image quality assessment through FSIM, SSIM, MSE and PSNR—a comparative study. Journal of Computer and Communications, 7(3), 8-18.

*Table 7. Features obtained for the dataset and their descriptions.*

| | Feature | Description |
|---|---|---|
| PING | average_rtt | RTT, or Round Trip Time, is known as the time it takes for a computer to send a packet to a destination and for the destination to receive this packet and respond. 'average_rtt' refers to the average of RTT values occurring within a specific period of time. |
| | packet_loss | It refers to the situation where packets sent over the network fail to reach their destination. It indicates the percentage of packet loss. |
| IPERF3 | bits_per_second (throughput) | It indicates the number of bits transmitted or received each second. |
| | bu_ratio | It indicates the instantaneous bandwidth utilization rate on a specific path. |
| | retransmits | It refers to the number of packets retransmitted by the sender in case of packet loss or incomplete delivery during data transmission. |
| | cpu_host_total | It indicates the total CPU usage rate on the host. |
| | cpu_host_user | It indicates the portion of CPU usage related to user-executed processes on the host. |
| | cpu_host_system | It indicates the portion of CPU usage related to system-level processes on the host. |
| | cpu_remote_total | It indicates the total CPU usage on a server accessed over a network. |
| | cpu_remote_user | It indicates the portion of CPU usage related to user-executed processes on a server accessed over a network. |
| | cpu_remote_system | It indicates the portion of CPU usage related to system-level processes on a server accessed over a network. |
| NFSTREAM | bidirectional_duration_ms | It expresses the round-trip time in a communication link, which refers to the time it takes for a signal to travel from the sender to the receiver and back again. |
| | bidirectional_packets | It refers to the data packets transmitted in both directions of a communication link. |
| | bidirectional_bytes | It refers to the total amount of bytes transferred in both directions in a communication link. |
| | src2dst_duration_ms | It refers to the duration of data transferred from a source to a destination on a network, typically expressed in milliseconds. |
| | src2dst_packets | It refers to the number of packets sent during data transfer from a source to a destination on a network. |
| | src2dst_bytes | It refers to the total number of bytes sent during data transfer from a source to a destination on a network. |
| | dst2src_duration_ms | It refers to the duration of data transfer from a destination to a source on a network, typically expressed in milliseconds. |
| | dst2src_packets | It refers to the number of packets sent during data transfer from a destination to a source on a network. |
| | dst2src_bytes | It refers to the total number of bytes sent during data transfer from a destination to a source on a network. |
| | bidirectional_min_ps | It refers to the minimum packet size of data transferred in both directions in a communication link. |
| | bidirectional_mean_ps | It refers to the average packet size of data transferred in both directions in a communication link. |
| | bidirectional_stddev_ps | It refers to the standard deviation of packet size in data transfer in both directions in a communication link. |
| | bidirectional_max_ps | It refers to the maximum packet size of data transferred in both directions in a communication link. |
| | src2dst_min_ps | It refers to the minimum packet size of data transferred from a source to a destination in a communication link. |
| | src2dst_mean_ps | It refers to the average packet size of data transferred from a source to a destination in a communication link. |
| | src2dst_stddev_ps | It refers to the standard deviation of packet size in data transfer from a source to a destination in a communication link. |
| | src2dst_max_ps | It refers to the maximum packet size of data transferred from a source to a destination in a communication link. |
| | dst2src_min_ps | It refers to the minimum packet size of data transferred from a destination to a source in a communication link. |
| | dst2src_mean_ps | It refers to the average packet size of data transferred from a destination to a source in a communication link. |
| | dst2src_stddev_ps | It refers to the standard deviation of packet size in data transferred from a destination to a source in a communication link. |
| | dst2src_max_ps | It refers to the maximum packet size of data transferred from a destination to a source in a communication link. |
| | bidirectional_min_piat_ms | It refers to the minimum packet arrival time of data transferred in both directions in a communication link. |
| | bidirectional_mean_piat_ms | It refers to the average packet arrival time of data transferred in both directions in a communication link. |
| | bidirectional_stddev_piat_ms | It refers to the standard deviation of packet arrival times in data transferred in both directions in a communication link. |
| | bidirectional_max_piat_ms | It refers to the maximum packet arrival time of data transferred in both directions in a communication link. |
| | src2dst_min_piat_ms | It refers to the minimum packet arrival time of data transferred from a source to a destination in a communication link. |
| | src2dst_mean_piat_ms | It refers to the average packet arrival time of data transferred from a source to a destination in a communication link. |
| | src2dst_stddev_piat_ms | It refers to the standard deviation of packet arrival times in data transferred from a source to a destination in a communication link. |
| | src2dst_max_piat_ms | It refers to the maximum packet arrival time of data transferred from a source to a destination in a communication link. |
| | dst2src_min_piat_ms | It refers to the minimum packet arrival time of data transferred from a destination to a source in a communication link. |
| | dst2src_mean_piat_ms | It refers to the average packet arrival time of data transferred from a destination to a source in a communication link. |
| | dst2src_stddev_piat_ms | It refers to the standard deviation of packet arrival times in data transferred from a destination to a source in a communication link. |
| | dst2src_max_piat_ms | It refers to the maximum packet arrival time of data transferred from a destination to a source in a communication link. |
| | psnr | It is a metric that measures the quality of an image. It compares the transmitted or compressed image with the original image to evaluate packet loss that may occur during image compression or transmission. |
| | ssim | It serves to measure the quality of an image. It is a more suitable measure for the human eye compared to PSNR and focuses on structural similarity. |
| LINUX COMMAND | original_file_size | It refers to the original file size. |
| | file_size | It refers to the file size of the transmitted file. |
| FLOODLIGHT | hop_count | It indicates the number of intermediate devices (routers or switches) that a data packet must pass through to reach its destination. |